\documentclass[twocolumn]{emulateapj}

\usepackage{amsmath}
\usepackage{txfonts}
\usepackage{latexsym}
\usepackage{amsfonts}  
\usepackage{amssymb}
\usepackage{amsbsy}   
\usepackage{natbib}
\usepackage[english]{babel}

\newcommand{\dt}[1]{\frac{\partial #1}{\partial t}}

\newcommand{\dr}[1]{\frac{\partial #1}{\partial r}}
\newcommand{\ddr}[1]{\frac{\mathrm{d}#1}{\mathrm{d}r}}

\newcommand{\age}{\gtrsim}
\newcommand{\ale}{\lesssim}

\newcommand{\ba}{\begin{eqnarray}}
\newcommand{\ea}{\end{eqnarray}}
\newcommand{\bas}{\begin{eqnarray*}}
\newcommand{\eas}{\end{eqnarray*}}
\newcommand{\be}{\begin{equation}}
\newcommand{\ee}{\end{equation}}
\newcommand{\bes}{\begin{equation*}}
\newcommand{\ees}{\end{equation*}}
\newcommand{\bfd}{\begin{figure}[tbh]}
\newcommand{\bft}{\begin{figure}[t]}
\newcommand{\bfh}{\begin{figure}[h]}
\newcommand{\bfb}{\begin{figure}[b]}
\newcommand{\ef}{\end{figure}}
\newcommand{\bd}{\begin{displaymath}}
\newcommand{\ed}{\end{displaymath}}

\newcommand{\elsasser}{Els\"{a}sser }
\newcommand{\alfven}{Alfv\'en }

\newcommand{\alfvenic}{Alfv\'enic }

\shorttitle{Reflection-driven turbulence}
\shortauthors{Verdini et al.}

\begin{document}

\title{Turbulence in the sub-Alfv\'enic solar wind driven by reflection of 
low-frequency Alfv\'en waves}

\author{A. Verdini\altaffilmark{1}, M. Velli\altaffilmark{2,3} 
and E. Buchlin\altaffilmark{4}}
\altaffiltext{1}{Observatoire Royale de Belgique, 3 Avenue Circulaire, 1180,
Bruxelles, Belgium; e-mail: verdini@oma.be}
\altaffiltext{2}{Dipartimento di Astronomia e Scienza dello Spazio, Univ. di Firenze,
Largo E. Fermi 3, 50125, Firenze, Italy}
\altaffiltext{3}{Jet Propulsion Laboratory, California Institute of Technology, 4800 Oak Grove Drive, Pasadena, CA 91109, USA}
\altaffiltext{4}{Institut d'Astrophysique Spatiale, CNRS - Universit\'e
Paris Sud, B\^at. 121, 91405, Orsay Cedex, France}

\begin{abstract}
We study the formation and evolution of a turbulent
spectrum of \alfven waves driven by reflection off the solar wind density
gradients, starting from the coronal base up to 17
solar radii, well beyond the \alfvenic critical point. The background solar
wind is assigned and 2D shell models are used to describe nonlinear interactions.
We find that the turbulent spectra are influenced by the nature of reflected waves. Close to the base, these give rise to a flatter and steeper spectrum for 
the outgoing and reflected waves respectively. 
At higher heliocentric distance both spectra evolve toward an asymptotic Kolmogorov spectrum. 
The turbulent dissipation is found to account for at least half of the
heating required to sustain the background imposed solar wind and its shape is
found to be determined by the reflection-determined turbulent heating below 1.5 solar radii. 
Therefore reflection and reflection-driven turbulence are shown to play a key role in the acceleration 
of the fast solar wind and origin of the turbulent spectrum found at 0.3~AU in the heliosphere.
\end{abstract}

\keywords{MHD --- waves --- turbulence --- solar wind}

\section{Introduction}
Recent high resolution observations from Hinode \citep{DePontieu_al_2007} and
the Coronal Multi-channel Polarimeter \citep{Tomczyk_al_2007} 
have shown that the
solar atmosphere is pervaded by Alfv\'enic (or kink-like, e.g. \citealp{VanDoorsselaere_al_2008}) oscillations: 
observed in jets, spicules or in coronal loops,  velocity and magnetic
field oscillations ($\delta b,~\delta u$) are coupled and propagate at speeds close to the
\alfven speed. 
Far from the sun, between 0.3~AU and several AU, in situ data show that in
the frequency range $10^{-4}~\mathrm{Hz}\ale f\ale 10^{-2}~\mathrm{Hz}$, fluctuations
in magnetc field and velocity $\delta b$ and $\delta u$ possess many of the properties of
\emph{outward-propagating} "spherically polarized"  \alfven waves, namely: quasi incompressibility, 
correlated oscillations, and a constant (total) magnetic field intensity, while at the same time revealing
their turbulent nature through a well-developed  power-law frequency spectrum, with a break separating
different power law slopes of -1 and -1.6 which moves to lower frequencies with 
increasing distance from the sun \citep{Bavassano_al_1982,Tu_al_1984}.
The $\delta u\cdot\delta b$ correlation, upon which the propagation direction determination is made, 
depends on the frequency considered and varies with distance \citep{Bavassano_al_2000a,Bavassano_al_2000b} and latitude
\citep{Grappin_2002}, typically in the range 
$1/2<|(\delta u\cdot\delta b/\sqrt{4 \pi\rho})/(\delta u^2 + \delta
b^2/4\pi\rho)|<1$
(perfect correlation corresponding to $1$). These facts suggest that the \emph{inward travelling} wave-mode component,
required for nonlinear couplings between incompressible fluctuations, must indeed be present.\\
This component might be generated locally between 0.3~AU and 1~AU by shear,
compressible or pick-up ions interactions, or it could be already 
formed in the sub-\alfvenic corona and later on nonlinearly advected into the
heliosphere by the solar wind, the hypothesis we consider here.

The dynamics inside the Alfv\'enic point region is of primary importance to understand
the origin of the spectrum one finds at 0.3~AU and whether 
it has any role in accelerating the solar wind. 
The variation of the propagation speed induced by density
gradients in the stratified corona and accelerating solar wind causes 
outward Alfv\'en waves to be reflected, predominantly at lower frequencies,
hence triggering the incompressible cascade.
The power dissipated by the cascade contributes
to coronal heating, also modifying the overall turbulent pressure gradient,  fundamental to the acceleration of the fast solar wind.\\

While there are several studies on the \emph{linear} propagation and reflection 
of \alfven waves in the sub-\alfvenic corona and solar wind
\citep{Hollweg_1978a,Heinemann_Olbert_1980,
Velli_1993,Hollweg_Isenberg_2007}, 
excepting phenomenological models with an essentially dimensional estimate of the role of turbulent heating
 \citep{Hollweg_al_1982a,Dmitruk_al_2001a,
Cranmer_Ballegooijen_2005,Verdini_Velli_2007} 
very few of them have considered \emph{nonlinear} interactions.
\citet{Velli_al_1989, Velli_al_1990} studied the turbulent cascade
sustained by reflected waves in the super-\alfvenic solar wind, while
\citet{Dmitruk_al_2002} considered the same mechanism in the 
sub-\alfvenic corona below $3~R_\odot$, hence neglecting the solar wind.
\par
In the present letter we will extend these studies following the development of the turbulent cascade from the base of the corona up to $17~R_\odot$, well beyond
the \alfvenic critical point (located at about $13~R_\odot$ in the solar
wind model adopted). 
Direct numerical simulations are still prohibitively costly in terms of computational times, 
so nonlinear interactions are simulated using a 2D shell model 
\citep{Buchlin_Velli_2007}
which simplifies nonlinear interactions but still allows  4 decades in the perpendicular wavenumber space to be covered
while rigorously treating the propagation and reflection of waves along the radial mean magnetic field.
\section{Model description}
The equations describing the propagation of \alfven waves in an
inhomogeneous medium are derived from magnetohydrodynamics (MHD),
assuming that the large scale fields are stationary and separating
the time-fluctuating fields from the large-scale averages  \citep{Heinemann_Olbert_1980, Velli_1993}.
Therefore the large scale magnetic field, 
bulk wind flow and density ($B,~U,~\rho$ respectively) appear as specified
coefficients in the MHD equations for the fluctuations. 

We consider a magnetic
flux tube centered in a polar coronal hole, which expands 
super-radially with a (non-dimensional) area  $A(r)=r^2f(r)$ 
first prescribed by \citep{Kopp_Holzer_1976,Munro_Jackson_1977}. Distances are normalized to the solar radius, and the 
coefficients are given by $r_0=1.31,~\sigma=0.5,~f_{max}=7.26$ (respectively the location,
width, and asymptotic value of the super-radial expansion) so that A(1) =1.
The field becomes $B(r)=B_\odot /A(r)$ where we take $B_\odot=10~\mathrm{G}$.
The wind speed and density $U(r),~\rho(r)$ are obtained solving the 1D momentum equation
with an assigned temperature $T(r),~T_\odot=4~10^5~\mathrm{K}$ and a numerical
density at the coronal base 
$n_\odot=5~10^7~\mathrm{cm^{-3}}$
(see \citealp{Verdini_Velli_2007} and
references therein for details on the equation and on the temperature profile).
The resulting wind is supersonic far from the sun with
$U_{1\mathrm{AU}}\approx750~\mathrm{km/s}$ 
and 
$n_{1\mathrm{AU}}\approx3~\mathrm{cm^{-3}}$.
The \alfven critical point ($r_a$) is at about
$13~R_\odot$, the sonic critical point is at about $1.9~R_\odot$, 
the \alfven speed $V_a=B/\sqrt{4\pi\rho}$ at the base is $V_{a,\odot}\approx
1000~\mathrm{km~s^{-1}}$ and has a maximum
$V_{a,{max}}\approx3500~\mathrm{km/s}$ at $r=1.5~R_\odot$.
At the end of the domain ($17~R_\odot$) $U\approx740~\mathrm{km/s}$ and 
$V_a\approx630~\mathrm{km/s}$.\\
We assume $\delta u$ to be incompressible and
transverse with respect to $B$.
The momentum and induction equations for $\delta u$ and  $\delta b$ are
written in terms of the \elsasser fields $Z^\pm=\delta u \mp \delta
b/\sqrt{4\pi\rho}$, corresponding to \alfven waves which
propagate respectively outwards and inwards in the solar wind reference frame.

Substituting the nonlinear terms, which act in planes perpendicular to the radial direction, with a 2D MHD shell model representation (in the form
given by \citealp{Biskamp_1994})
and assuming radial propagation finally yields the model equations for
$Z^\pm=Z^\pm(r,k_\bot)=Z^\pm_n(r)$:
\ba
\dt{Z_n^\pm}+\left(U\pm V_a\right)\dr{Z_n^\pm}
+\frac{1}{2}\left(U\mp V_a\right)
\left(\ddr{\log V_a}+\ddr{\log A}\right)Z_n^\pm\nonumber\\
-\frac{1}{2}\left(U\mp V_a\right)
\left(\ddr{\log V_a}\right)Z_n^\mp=
-k_n^2\left(\nu^+Z_n^\pm+\nu^-Z_n^\mp\right)+ik_n(T_n^\pm)^*.
\label{eq:model}\ea
The complex scalar values, $u_n=(Z^+_n+Z^-_n)/2$,
$b_n=(Z^-_n-Z^+_n)/2$, 
represent the velocity and magnetic (in velocity units) field fluctuations corresponding to the scale
$\lambda_n=\lambda_02^{-n}=2\pi/k_n$, $n$ is the shell index,
and $T_n^\pm$ accounts for nonlinear interactions of the form 
$Z^+_l Z^-_m$, with $l,m=n\pm1,n\pm2$.
Finally, $\nu^\pm=(\nu\pm\eta)/2$ are combinations of the kinematic viscosity and
the magnetic resistivity (we take $\eta=\nu$).\\
Simulations are carried out with the code \textsc{Shell-Atm}\footnote{
modified to include the wind spherical expansion} \citep{Buchlin_Velli_2007}.
The advection terms in
eq.~\ref{eq:model} are computed with a second order upwind scheme 
(Fromm scheme) which
allows a good conservation of the phase of the fluctuation. Time is advanced
with a third order Runge-Kutta for the nonlinear 
part of the equations.
The radial domain is decomposed in $\sim25,000$ planes over a non-uniform grid 
while 21 shells are used for the
nonlinear interactions.
Transparent boundary conditions are imposed at the top for both waves, and at the
bottom for the $Z^-$. Here all the gradients are artificially set to zero, in order to avoid reflections.
Energy is injected in the domain imposing
the amplitude $Z^+_{n,\odot}=f_n(t)$ at the first 3 shells corresponding to length
scale of the order of 8.000-34.000~km ($\lambda_{0,\odot}=0.02~R_\odot$ in the shell model), with $f(t)$ a function with a
time correlation and periodicity $\tau^*\approx 1000~\mathrm{s}$. The form of the
function is given in 
\citet{Buchlin_Velli_2007}, here it is important to note that despite the
correlation time is only 1000~s, some \emph{low-frequency} fluctuations are injected for long time series (as in the present simulation). 
Simulations last about 20 crossing time scale, 
$\tau_{cr}=\int^{17R\odot}_{R\odot}\approx7000~\mathrm{s}$, and time averaged
quantities are computed on the last $10~\tau_{cr}$, during which the system has an approximate stationary state.
\begin{figure*}
\plotone{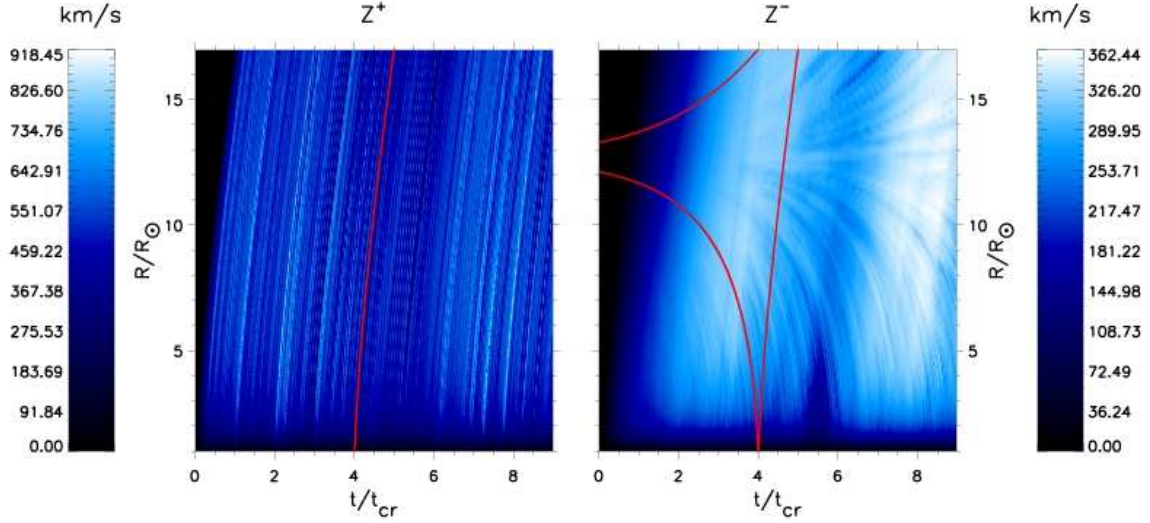}
\caption{Root mean squared amplitudes of the mother wave $Z^+$ and the
reflected waves $Z^-$ in $\mathrm{km~s^{-1}}$ (left and right panels respectively)
as a function of heliocentric distance and time (in unit of
$\tau_{cr}$),
along with the outgoing and ingoing characteristics $U\pm V_a$ (red curves).
Two different paths of the $Z^-$ can be distinguished in the right panel:
one associated to its phase speed ($U-V_a$, negative below
the \alfvenic critical point at $\approx13~R_\odot$)
and the other associated to reflection from the $Z^+$ which forces the
fluctuations to follow the outgoing wave (with phase speed $U+V_a$).}
\label{fig:1}
\end{figure*}
\section{Reflection and nonlinearities}
Given that we start with an initial flux
of $Z^+$ at the coronal base, reflection is the only trigger 
for nonlinear interactions.
In Figure~\ref{fig:1} the root mean squared (rms) amplitudes
$|Z^\pm|=\sqrt{\sum_n |Z_n^\pm|^2}$
are plotted as a function of time and heliocentric distance.
Two components of the reflected waves are clearly visible in the $Z^-$ contours:
a ``classical'' component $Z^-_{class}$, which propagate with the expected $U-V_a$ phase speed
(negative below the \alfvenic critical point, $r_a\approx13~R_\odot$) and an
``anomalous'' component $Z^-_{anom}$, which travels with the same speed as the mother wave, $U+V_a$.
As shown analytically and numerically
\citep{Velli_al_1990,Hollweg_Isenberg_2007} 
this ``anomalous'' component is the direct
product of reflection.
Generally in presence of density gradients, 
for small values of the ratio $\epsilon=\alpha/\omega$, 
typical of the upper corona, with
$\alpha=\tau_R^{-1}=|(U\mp V_a)V_a'/V_a|$ in eq.~\ref{eq:model}, 
each field can be decomposed in a primary and secondary
component. A $Z^+$ primary component is injected at the base, while
the $Z^-$ is made up of only the secondary component, given by reflection.
In each plane, as the $Z^+$ arrives, the secondary component can be seen as
the result of a forcing term given by $\approx\alpha Z^+$, hence producing a
wave which travels with the same phase speed and of the above amplitude.
The value $Z^- = \epsilon Z^+$ follows naturally by finding the ``forced'' solution to the linearized equation for  $Z^- $.
At later time, as the $Z^+$ has propagated away,
the forcing disappears and the secondary component propagates 
backward with the classical phase speed. When a $Z^+$ pulse wave is excited in
the corona, $Z^-$ appears as a halo spreading backward from
the mother $Z^+$ wave.
Nonlinear interactions modifies the above picture, 
acting as a local (in a given plane) source for the $Z_n^-$
which is uncorrelated with respect to that given by
reflection and hence generating waves propagating with classical phase speed
$U-V_a$, i.e., a primary component.\\
Inward propagating waves have a very long propagation time at $r_a$ 
that slows down the overall relaxation toward a steady state: for example in
Figure~\ref{fig:1} the increase of
the $Z^-$ amplitude at $t\approx 8\tau_{cr}$ results from the superposition of
the $Z^-_{anom}$ with the backward propagating $Z^-_{class}$ produced at
$t\approx3\tau_{cr}$.
\par
\begin{figure}
\plotone{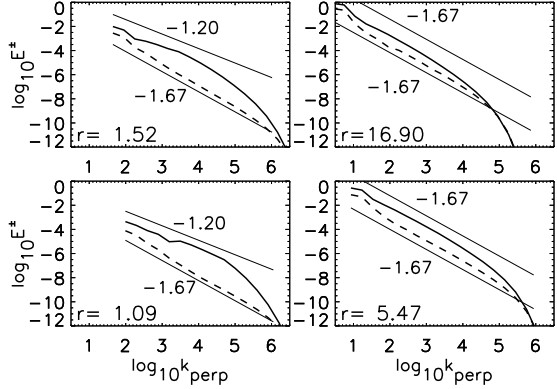}
\caption{Energy spectra $E_k^\pm$ (solid and dashed lines respectively)
as a function of the perpendicular wavenumber in
four different planes, as indicated at the bottom left of each
panel ($r=R/R_\odot$). Spectra are averaged in time and normalized 
to $\approx1.5\times10^{25}~\mathrm{cm^3~s^{-2}}$, wavenumbers to $R_\odot^{-1}$. 
Thin lines are power-law with the indicated slopes, plotted for reference.}
\label{fig:2}
\end{figure}
The different nature of the reflected waves influences the spectral energy
transfer. 
Generally speaking, while the wave amplitudes increase with distance
the nonlinear timescales $\tau_{nl}^\pm=(k_n Z^\mp_n)^{-1}$ decrease,
following the reflection coefficient and the flux tube expansion.
At small wavenumbers $\tau_{nl}^-< \tau_{nl}^+$, the
energy transfer is more efficient for the reflected waves, although
the $Z^+$'s contribute more to the total energy dissipation.
As a matter of fact, the $Z^-$'s develop an inertial range in the whole
domain and the resulting cascade is more ``aged'', 
their dynamics is governed by the nonlinear interactions which
have shorter timescales (as $Z^+> Z^-$) and act for longer periods, the $Z^-$
having a smaller propagation speed.\\
The contribution of coherent interactions $Z^+Z^-_{anom}$ 
in shaping the energy spectra
$E^\pm_k=|Z^\pm_n|^2/(4k_n)$ can be seen in the low corona, before
$1.2~R_\odot$ (left panels of Figure~\ref{fig:2}). 
In fact at $1.09~R_\odot$, assuming waves of period $T=1000~\mathrm{s}$,
$\epsilon\approx0.07$ ($\epsilon_{max}=0.16$ at the coronal
base), hence in presence of lower frequency fluctuations (injected by the
forcing) reflection is relatively high. 
At about $1.5~R_\odot$, $V_a$ has a maximum and reflection vanishes, so
that the $Z^-$ are not produced locally but rather propagating from above. This
inhibits the spectral energy flux of $Z^+$ that becomes negative at large
scales. 
The inverse spectral transfer is responsible of the flatter part of the
$E^+$ spectrum one finds at small $k_\perp$.
Further out reflection is negligible and the spectra are the remnant of those
produced in the inner layers, showing a slow evolution toward the asymptotic
state in which $E^+\age3E^-$. The spectral energy flux is given by the
interctions $Z^+Z^-_{class}$ that are subject to the ``\alfven
effect'' (a longer cascade timescale due to the nonlinear interactions between wavepackets propagating in opposite directions), producing the same spectral slopes -5/3 for $E^\pm$ 
(right panels of Figure~\ref{fig:2}).\\
Note that the largest
perpendicular scale is proportional to the flux tube
width, $\lambda_0(r)=\lambda_{0,\odot}\sqrt{A(r)}$, hence
the same number of shells spans a $k_\bot$ interval
that shifts to smaller values with increasing $r$.

The asymptotic slope $-5/3$ and that one of the reflection dominated spectra can be deduced from the expression of spectral energy flux of the shell model 
($a_1=11/24,~a_2=1-a_1,~a_3=-15/24$), 
\ba
\Pi^\pm_n &=&-\mathrm{Im}\left[k_nZ_n^+\left(a_1Z^+_{n+1}Z^-_{n+2} +
a_2Z^-_{n+1}Z^+_{n+2}\right)\right.  \nonumber\\
 & + & \left.k_{n-1}Z^+_{n+1}\left(a_2Z^+_{n-1}Z^-_{n} +
a_3Z^-_{n-1}Z^+_{n}\right)\right]
\label{eq:kflux}
\ea
When reflection is negligible one can assume that 
the $Z^+$ and $Z^-$ are uncorrelated $\langle Z^+{Z^-}^*\rangle_t\approx0$.
Assuming a power law for the spectral energies,
$E_{n}^\pm\propto(k_n/k_0)^{p_\pm}$, substituting 
$Z_n^\pm=2(k_nE_{n}^\pm)^{1/2}$ in eq.~\ref{eq:kflux}
one finds that $\Pi^\pm_n$ is independent of the shell index $n$ 
when $p^+=p^-=-5/3$. 
For a given plane, it implies a constant normalized cross helicity
$\sigma_c=(E^+-E^-)/(E^++E^-)$ 
in the inertial range, with a cross helicity spectrum $H_c=E^+-E^-$ of the same
slope as $E^+$, in contrast to EDQNM closure models, which, \emph{including
nonlocal interactions} in Fourier space, predict a distinct steeper spectrum, $H_c\propto k^{-2}$ \citep{Grappin_al_1983}.
From the evolution of the spectra (right panels) one can see that
the normalized cross helicity decreases with $r$ because of two factors, the general increase of the total energy and the decrease of
$H_c$ at all scales. The former is due to the approximate conservation of the
total wave action density.
The later results from the competition of
the linear coupling (reflection), which
forces $Z^-\ale Z^+$ for low frequency fluctuations, with the
nonlinear coupling, which damps the non dominant wave population, the $Z^-$
\citep{Dmitruk_al_2001a}.
Indeed the $E^\pm$ spectra look the same at small wavenumbers 
(corresponding to $n=0,~1,~2$) in which low frequency fluctuations reside.\\

When reflection is high the two fields are strongly coupled and one can
assume that the growth of $Z^-_n$ has contributions both fromt the nonlinear cascade and direct generation
through reflection $\approx\epsilon_n Z^+_n$. The reflected contribution shares the phase properties
of $Z^+_n$ so interactions due to reflection are coherent below
$1.2~R_\odot$ and the cascade may be isotropic
for small wavenumbers in the $k_\bot-k_{||}$ plane.
For the first 7 shells (2 orders of
magnitude), one would find  $\omega_n=k_nV_a$, $\epsilon_n\propto\lambda^{-n}$ 
and the following scaling for the spectra
$E^+\propto k^{-1}$, $E^-\propto k^{-3}$ \citep{Velli_al_1990}. 
If the nonlinear cascade is some how inhibited, keeping fixed the total
energy in the parallel wavenumbers, low frequencies are
confined to smaller $n$ and $\epsilon_n\propto
\lambda^{-np}$ with $p<1$, yielding a steeper (flatter) spectrum for $E^+$
($E^-$) compared to the isotropic case. In the present simulations,
$\epsilon_{1.09R_\odot}\propto \lambda^{0.7}$ for $n<8$ which yields $p^+=1.2$,
$p^-=2.6$ as in the bottom left panel of Figure~\ref{fig:2}.\\
\par
\begin{figure}
\plotone{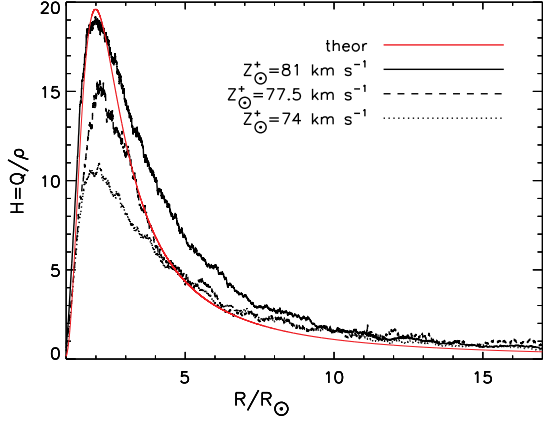}
\caption{Time averaged heating per unit mass as a function of distance for
different root mean squared amplitudes at the coronal base $Z^+_\odot$. Also
plotted in red is the heating function necessary to sustain the background
specified solar wind. $H$ is in units of $\approx 3\times
10^{10}\mathrm{cm^2~s^{-3}}$.}
\label{fig:3}
\end{figure}
The resulting turbulent heating is computed from eq.~\ref{eq:model}
multiplied by $(Z^\pm_n)^*$ and summed over the shell index. The total energy dissipation per unit mass
$H=Q/\rho=1/2\nu\sum_n k_n^2(|Z^+_n|^2 + |Z^-_n|^2)$
increases
in the low corona and decreases exponentially in the sub-\alfvenic region of
the wind, as expected.
For $\delta u_\odot\approx 49~\mathrm{km~s^{-1}}$ (solid line in
Figure~\ref{fig:3}), it is also 
very close to the heating required to
sustain the specified background wind (red line in Figure~\ref{fig:3}): the
location and height of the peak coincide, although more energy is dissipated in
the upper corona.
This would produce a faster wind, not altering too
much the mass flux, since the peak of the dissipation is close to 
the sonic critical point \citep{Hansteen_al_1999}.\\
For smaller values of $\delta u_\odot$ a better agreement is found in the
decreasing part, the rapid increase
in the low corona is still reproduced but the peak intensity is not attained.
Despite the fact that the spectra and the spectral fluxes
possess the same properties, 
for $\delta u_\odot\approx 45$~and~$42~\mathrm{km~s^{-1}}$ the
peak intensity decreases by a factor 1/4~and~1/2 respectively (dashed and dotted
line in Figure~\ref{fig:3}).
It turns out that the amount of dissipation is very sensitive to the level of
velocity fluctuations at the base of the corona.
In fact, $\delta u$ is determined by the injection of $Z^+$ at
the coronal base but also by the response of the atmosphere and by the nonlinear
interactions (the level of $Z^-$). On the countrary the peak dissipation seems
to scale linearly with the rms amplitude $Z^+_\odot$ although further studies
(dependence on the frequency, on the nonlinear interactions in the 
shell model, on the imposed wind) are necessary to define the scaling precisely.\\
\section{Discussion}
We have studied the propagation, reflection and nonlinear
interaction of \alfven waves from the base of the corona up to 17 solar
radii, well beyond the \alfvenic critical point.
For the first time 2D shell models have been applied to account for
nonlinear interactions in magnetically open regions on the sun, such as coronal holes. 
Thanks to such a simplification, compared to MHD or Reduced MHD direct numerical
simulations, it is possible to
follow the development of a turbulent spectrum in the expanding
solar wind, where waves are continuously reflected by the gradients in mean fields.
Reflected waves are made of two components, one propagating
with the characteristic phase speed $U-V_a$ ($Z^-_{class}$) and the other
following the path of the outgoing wave with speed $U+V_a$
($Z^-_{anom}$), a confirmation of previous linear results
(\citealp{Velli_al_1989,Hollweg_Isenberg_2007}) which
hold in a similar way also in the nonlinear regime. 
For typical coronal 
parameters we find $Z^-=\epsilon(r,\omega)Z^+\propto V_a'Z^+$, in contrast to
$Z^-\propto \lambda_0V_a'$
(independet of $Z^+$) found in a strong turbulence regime
\citep{Dmitruk_al_2002} in which $Z^-,~\lambda_0\rightarrow0$.
Differences arise because the above limits impose a time scale ordering  $\tau_{nl}^-<<\tau_{nl}^+\ale \tau_R<\tau_{cr}$ 
which is not satisfied in our simulation, basically because $Z^-\nrightarrow0$
even in case in which the outer scale of turbulence does not follow the flux
tube expansion (similar to the limit $\lambda_0\rightarrow0$).\\
Close to the coronal base one can distinguish the contribution to the spectral
slope of the coherent nonlinear interactions $Z^-_{anom}Z^+$, which give  
$E^+\propto k^{-1.2}$, and the one from the 
incoherent nonlinear interaction $Z^-_{class}Z^+$, giving $E^\pm\propto
k^{-5/3}$ (in this shell model, which includes nonlinear interactions
only locally in Fourier space). The resulting spectra change with
distance, starting from a coherent-interaction dominated spectrum at the
coronal base and evolving toward the asymptotic Kolmogorov spectra at greater
distance, where reflection is negligible. According to this model, outside the
\alfvenic critical point, the turbulent spectra have already lost any feature
acquired in the low corona. Note that this is referred to perpendicular
wavenumber spectra and not to the frequency spectra, which on the countrary
are almost unchanged, since their evolution is limited to the first solar
radius above the coronal base.\\
Turbulent dissipation is remarkably high in the low corona.
Depending on the injected energy an almost complete or partial matching is
found with the ``theoretical'' heating, that is the one required to form the
imposed background solar wind.
The best agreement is found for $\delta u_\odot\approx 50~\mathrm{km~s^{-1}}$ which is at the limit of observational
constraints \citep{Chae_al_1998}. Nonetheless, even for more conservative values
$\delta u_\odot\approx 40~\mathrm{km~s^{-1}}$, turbulent dissipation accounts 
for half of the above theoretical heating, maintaining the same profile (i.e. a
peak at the sonic point). This implies that the role of coherent interactions
is fundamental in shaping the heating function and that turbulence and
turbulent heating can not be neglected when studying the acceleration of the
solar wind.\\ The peak dissipation seems
to scale linearly with the rms amplitude $Z^+_\odot$ although a proof
of the precise scaling would require further studies.
We finally observe that if the lower boundary is shifted to the
base of the chromosphere, hence including the transition region,
a stronger dissipation rate is expected to be found at the transition
region and in the low corona. The \alfven speed is smaller below the transition region but its gradients are higher,
increasing the amount of energy residing in the anomalous reflected component which,
having more time to interact with its mother wave, might increase the
spectral energy transfer.\\
\par

{\it Acknowledgments.} This research was supported in part by 
ASI contract n.I/015/07/0 "Solar System Exploration" 
and it was carried out in part at JPL under a contract from NASA.
A.~Verdini acknowledges support from the Belgian Federal Science Policy Office through the ESA-PRODEX program.


\begin{thebibliography}{26}
\expandafter\ifx\csname natexlab\endcsname\relax\def\natexlab#1{#1}\fi

\bibitem[{Bavassano {et~al.}(1982)Bavassano, Dobrowolny, Mariani, \&
  Ness}]{Bavassano_al_1982}
Bavassano, B., Dobrowolny, M., Mariani, F., \& Ness, N.~F. 1982, J. Geophys.
  Res., 87, 3616

\bibitem[{{Bavassano} {et~al.}(2000{\natexlab{a}}){Bavassano}, {Pietropaolo},
  \& {Bruno}}]{Bavassano_al_2000b}
{Bavassano}, B., {Pietropaolo}, E., \& {Bruno}, R. 2000{\natexlab{a}}, J.
  Geophys. Res., { 105}, 12697

\bibitem[{{Bavassano} {et~al.}(2000{\natexlab{b}}){Bavassano}, {Pietropaolo},
  \& {Bruno}}]{Bavassano_al_2000a}
{Bavassano}, B., {Pietropaolo}, E., \& {Bruno}, R. 2000{\natexlab{b}}, J.
  Geophys. Res., { 105}, 15959

\bibitem[{{Biskamp}(1994)}]{Biskamp_1994}
{Biskamp}, D. 1994, Phys. Rev. E, { 50}, 2702

\bibitem[{{Buchlin} \& {Velli}(2007)}]{Buchlin_Velli_2007}
{Buchlin}, E. \& {Velli}, M. 2007, \apj, 662, 701

\bibitem[{{Chae} {et~al.}(1998){Chae}, {Yun}, \& {Poland}}]{Chae_al_1998}
{Chae}, J., {Yun}, H.~S., \& {Poland}, A.~I. 1998, ApjS, { 114}, 151

\bibitem[{{Cranmer} \& {van Ballegooijen}(2005)}]{Cranmer_Ballegooijen_2005}
{Cranmer}, S.~R. \& {van Ballegooijen}, A.~A. 2005, ApjS, { 156}, 265

\bibitem[{De~Pontieu {et~al.}(2007)De~Pontieu, McIntosh, Carlsson, Hansteen,
  Tarbell, Schrijver, Title, Shine, Tsuneta, Katsukawa, Ichimoto, Suematsu,
  Shimizu, \& Nagata}]{DePontieu_al_2007}
De~Pontieu, B., McIntosh, S.~W., Carlsson, M., {et~al.} 2007, Science, 318,
  1574

\bibitem[{{Dmitruk} {et~al.}(2002){Dmitruk}, {Matthaeus}, {Milano}, {Oughton},
  {Zank}, \& {Mullan}}]{Dmitruk_al_2002}
{Dmitruk}, P., {Matthaeus}, W.~H., {Milano}, L.~J., {et~al.} 2002, ApJ, { 575},
  571

\bibitem[{{Dmitruk} {et~al.}(2001){Dmitruk}, {Milano}, \&
  {Matthaeus}}]{Dmitruk_al_2001a}
{Dmitruk}, P., {Milano}, L.~J., \& {Matthaeus}, W.~H. 2001, ApJ, { 548}, 482

\bibitem[{{Grappin}(2002)}]{Grappin_2002}
{Grappin}, R. 2002, Journal of Geophysical Research (Space Physics), 107, 1247

\bibitem[{{Grappin} {et~al.}(1983){Grappin}, {Leorat}, \&
  {Pouquet}}]{Grappin_al_1983}
{Grappin}, R., {Leorat}, J., \& {Pouquet}, A. 1983, Astron.\ Astrophys., 126,
  51

\bibitem[{{Hansteen} {et~al.}(1999){Hansteen}, {Leer}, \&
  {Lie-Svendsen}}]{Hansteen_al_1999}
{Hansteen}, V.~H., {Leer}, E., \& {Lie-Svendsen}, {\O}. 1999, in ESA SP-448:
  Magnetic Fields and Solar Processes, ed. A.~{Wilson} \& {et al.}, 1091--+

\bibitem[{{Heinemann} \& {Olbert}(1980)}]{Heinemann_Olbert_1980}
{Heinemann}, M. \& {Olbert}, S. 1980, J. Geophys. Res., { 85}, 1311

\bibitem[{{Hollweg}(1978)}]{Hollweg_1978a}
{Hollweg}, J.~V. 1978, \solphys, { 56}, 305

\bibitem[{{Hollweg} \& {Isenberg}(2007)}]{Hollweg_Isenberg_2007}
{Hollweg}, J.~V. \& {Isenberg}, P.~A. 2007, Journal of Geophysical Research
  (Space Physics), 112, 8102

\bibitem[{{Hollweg} {et~al.}(1982){Hollweg}, {Jackson}, \&
  {Galloway}}]{Hollweg_al_1982a}
{Hollweg}, J.~V., {Jackson}, S., \& {Galloway}, D. 1982, \solphys, { 75}, 35

\bibitem[{{Kopp} \& {Holzer}(1976)}]{Kopp_Holzer_1976}
{Kopp}, R.~A. \& {Holzer}, T.~E. 1976, \solphys, { 49}, 43

\bibitem[{{Munro} \& {Jackson}(1977)}]{Munro_Jackson_1977}
{Munro}, R.~H. \& {Jackson}, B.~V. 1977, ApJ, { 213}, 874

\bibitem[{{Tomczyk} {et~al.}(2007){Tomczyk}, {McIntosh}, {Keil}, {Judge},
  {Schad}, {Seeley}, \& {Edmondson}}]{Tomczyk_al_2007}
{Tomczyk}, S., {McIntosh}, S.~W., {Keil}, S.~L., {et~al.} 2007, Science, 317,
  1192

\bibitem[{{Tu} {et~al.}(1984){Tu}, {Pu}, \& {Wei}}]{Tu_al_1984}
{Tu}, C.-Y., {Pu}, Z.-Y., \& {Wei}, F.-S. 1984, J. Geophys. Res., { 89}, 9695

\bibitem[{{Van Doorsselaere} {et~al.}(2008){Van Doorsselaere}, {Nakariakov}, \&
  {Verwichte}}]{VanDoorsselaere_al_2008}
{Van Doorsselaere}, T., {Nakariakov}, V.~M., \& {Verwichte}, E. 2008, ApJ
  Lett., 676, L73

\bibitem[{{Velli}(1993)}]{Velli_1993}
{Velli}, M. 1993, Astron.\ Astrophys., { 270}, 304

\bibitem[{{Velli} {et~al.}(1989){Velli}, {Grappin}, \&
  {Mangeney}}]{Velli_al_1989}
{Velli}, M., {Grappin}, R., \& {Mangeney}, A. 1989, Physical Review Letters, {
  63}, 1807

\bibitem[{{Velli} {et~al.}(1990){Velli}, {Grappin}, \&
  {Mangeney}}]{Velli_al_1990}
{Velli}, M., {Grappin}, R., \& {Mangeney}, A. 1990, Computer Physics
  Communications, 59, 153

\bibitem[{{Verdini} \& {Velli}(2007)}]{Verdini_Velli_2007}
{Verdini}, A. \& {Velli}, M. 2007, \apj, 662, 669

\end{thebibliography}
\end{document}